\documentclass[
preprint,
superscriptaddress,
showkeys,
 amsmath,amssymb,
 aps,
prc
]{revtex4-2}

\usepackage{graphicx}
\usepackage{dcolumn}
\usepackage{bm}
\usepackage{booktabs}
\usepackage{braket}
\usepackage[colorlinks]{hyperref}
\hypersetup{linkcolor=blue,citecolor=blue,urlcolor=blue}
\usepackage{makecell}

\usepackage{CJKutf8}
\usepackage{easyReview}

\newcommand{\Hg}{$ {^{180}\rm{Hg}} $ }

\newcommand{\ZnChan}{$ {^{68}\rm{Zn}} + {^{112}\rm{Sn}} $ }
\newcommand{\SeChan}{$ {^{74}\rm{Se}} + {^{106}\rm{Pd}} $ }
\newcommand{\KrChan}{$ {^{80}\rm{Kr}} + {^{100}\rm{Ru}} $ }
\newcommand{\KrChanb}{$ {^{84}\rm{Kr}} + {^{96}\rm{Ru}} $ }

\begin{document}
\begin{CJK*}{UTF8}{gbsn} 

\title{Shell effects in quasifission toward $^{180} \text{Hg}$: Insights into fission asymmetric modes}


\author{Yingge Huang (黄英格)} 
\affiliation{Sino-French Institute of Nuclear Engineering and Technology, Sun Yat-sen University, Zhuhai 519082, China}
\affiliation{Department of Physics, Graduate School of Science, The University of Tokyo, Tokyo 113-0033, Japan}

\author{Haozhao Liang (梁豪兆)} \email{haozhao.liang@phys.s.u-tokyo.ac.jp}
\affiliation{Department of Physics, Graduate School of Science, The University of Tokyo, Tokyo 113-0033, Japan}
\affiliation{Quark Nuclear Science Institute, The University of Tokyo, Tokyo 113-0033, Japan}
\affiliation{RIKEN Center for Interdisciplinary Theoretical and Mathematical Sciences (iTHEMS), Wako 351-0198, Japan}

\author{Jun Su (苏军)}\email{sujun3@mail.sysu.edu.cn} 
\affiliation{Sino-French Institute of Nuclear Engineering and Technology, Sun Yat-sen University, Zhuhai 519082, China}
\affiliation{Key Laboratory of Nuclear Data, China Institute of Atomic Energy, Beijing 102413, China}%

\date{\today}

\begin{abstract}
\begin{description}
\item[Background]
The experiment of $^{180}\rm{Hg}$ fission revealed a possible ``new asymmetric fission mode'' in the preactinide region, posing challenges to current fission theory.
Similarity in shell effects is observed between fission and quasifission, providing the possibility for widely exploring the topography of a fission potential-energy surface (PES).

\item[Purpose]
We aim to investigate the shell effects in the quasifission toward $^{180} \rm{Hg}$ and to explore their connection with the $^{180} \rm{Hg}$ fission.

\item[Method]
$^{68}\rm{Zn}+{^{112}\rm{Sn}}$, $^{74}\rm{Se}+{^{106}\rm{Pd}}$, $^{80}\rm{Kr}+{^{100}\rm{Ru}}$, and $^{84}\rm{Kr}+{^{96}\rm{Ru}}$ central collisions at different energies and projectile orientations are calculated using the Skyrme time-dependent Hartree-Fock approach.
The static fission properties are calculated with the constrained Hartree-Fock-Bogoliubov method and compared with the quasifission results.

\item[Results]
Shell effects are found to hinder mass equilibration between the prefragments, enhancing the production of fragments near the $80/100$ mass split.
By comparing the quasifission trajectories with the PES in the $(Q_{20}, Q_{30})$ space, 
the role of the PES ridge in forming fragments is identified.
The presence of the asymmetric valley causes the $ {^{68}\rm{Zn}} + {^{112}\rm{Sn}} $ quasifission to exhibit a prefragment mass equilibration process and scission-point configuration similar to those of $^{180} \rm{Hg}$ fission.
The elongated light fragment is found to be a key factor in reproducing the experimental fission total kinetic energies.
Meanwhile, a more pronounced proton shell gap is found in the $ {^{68}\rm{Zn}} + {^{112}\rm{Sn}} $ quasifission compared with other reactions.

\item[Conclusions]
By using quasifission dynamics as a probe of the fission pathway, the present calculations help clarify the specific influence of the PES topography and provide support for the dominance of proton shell effects and light fragment deformation in preactinide fission.

\end{description}
\end{abstract}

\maketitle
\end{CJK*}
\section{\label{sec:intro}Introduction}

Nuclear fission is a complex quantum many-body process that involves large-amplitude collective motion. 
Substantial progress has been made in understanding the phenomena, yet further investigation remains necessary, which is also important in studies of the nuclear industry, $r$-process nucleosynthesis and superheavy elements synthesis \cite{Andreyev2018RPP,Schmidt2018RPP,Bender2020JPG,Bulgac2020Front,Schunck2022PPNP}. 
A crucial challenge is the role of shell effects. 
Similar to atomic systems, nuclei also exhibit shell structures that provide extra stability for certain proton or neutron numbers. 
When the nucleus is driven toward scission, these shell effects can significantly alter the reaction evolution, guiding the system toward particular configurations and influencing the final properties of the fragments.

In the actinide region, asymmetric fission can usually be explained reasonably well because of the strong shell effects.
However, this does not necessarily mean that the essential mechanisms are fully understood \cite{Moller2012PRC,Scamps2018Nature}.
The unexpected observation of asymmetric fission in \Hg from $\beta$-delayed fission experiments challenged the empirical expectation of predominantly symmetric splits \cite{Andreyev2010PRL,Elseviers2013PRC}.
While the following theoretical calculations can reproduce fragment mass distributions, their physical interpretations remain under debate.
Within the macroscopic–microscopic framework, the mechanism is considered to differ from that in the actinides: the asymmetry is not driven by prefragment shell effects but instead by the hindrance of a potential–energy ridge.
This behavior is therefore referred to as “new asymmetric fission mode” \cite{Moller2012PRC,Ichikawa2012PRC,Ivanyuk2025PRC}.
In contrast, microscopic approaches indicate a dominance of shell effects associated with the neutron shell of heavy fragment  \cite{Scamps2019PRC,Bernard2023EPJA,Bernard2024EPJA}, 
which is also confirmed by the scission-point model \cite{Panebianco2012PRC,Huo2024EPJA}. 
Moreover, the current theoretical explanation remains insufficient to describe the systematic proton-number stabilization of a light fragment observed in preactinide fission \cite{Mahata2022PLB,Buete2025PLB,Morfouace2025Nature}.
These open issues make the preactinide region a crucial testing ground for investigating the manifestation of shell effects in fission.

Quasifission, a heavy-ion reaction in which the colliding nuclei undergo deep contact but do not form an equilibrated compound nucleus, is regarded as a possible means to populate fission modes \cite{Simenel2021PLB}.
Although quasifission and fission are distinct reaction processes, both involve relatively slow, strongly damped motion and comparable contact times between the prefragments  \cite{Toke1985quasi,Rietz2013Mapping}.
This allows the system to approach a near-equilibrium state even at large deformations, in which shell structure can exert a significant influence on the eventual splitting and can be interpreted in terms of the potential-energy surface (PES) \cite{Wakhle2014PRL,Morjean2017PRL,Hinde2022PRC,Pal2024PRC,McGlynn2023PRC,Lee2024PRC,Simenel2025PLB}.
In the preactinide region, the influence of the $Z=50$ shell has been considered as a factor enhancing the quasifission probability in reactions forming the \Hg compound nucleus \cite{Kozulin2021PLB,Itkis2025JPCS}.
However, the role of shell effects associated with the $80/100$ mass split in \Hg fission remains unclear.

To investigate the shell effects of quasifission forming preactinide nuclei, and their possible connection with fission,
we perform a comparative study of these processes in systems forming the \Hg compound nucleus, using both microscopic static and dynamical approaches.
This paper is organized as follows.
In Sec.~\ref{sec:model}, the methods for the static fission calculations and the quasifission  dynamics calculations are described.
In Sec.~\ref{sec:results}, the calculated results as well as some discussions are given.
In Sec.~\ref{sec:summary}, a summary of our work is presented.

\section{\label{sec:model}Theoretical framework and numerical calculations}

\subsection{HFB calculations of the fission potential-energy surface and path}

The nuclear quadrupole moment $Q_{20}$ and octupole moment $Q_{30}$ are typically used to describe the elongation and pear-shaped deformation of the system, respectively.
The fission PES is calculated from constrained Hartree-Fock-Bogoliubov (HFB) calculations with constraints on the \( Q_{20} \) and \( Q_{30} \). 
The constraints are imposed by fixing the expectation values \( \langle \hat{Q}_{20} \rangle = Q_{20} \) and \( \langle \hat{Q}_{30} \rangle = Q_{30} \) during the iterative energy minimization. 
The multipole operators are defined in the intrinsic frame as
\begin{align}
\hat{Q}_{20} &= \sqrt{\frac{5}{16\pi}} \left(2z^2 - r^2 \right), \\
\hat{Q}_{30} &= \sqrt{\frac{7}{16\pi}} \left[z(2z^2 - 3r^2) \right].
\end{align}

In addition, a one-dimensional fission path is calculated by constraining only the \( Q_{20} \). 
This provides the minimum-energy fission path, which approximately corresponds to the static adiabatic trajectory on the multidimensional PES. 
The comparison between the one-dimensional (1D) \( Q_{20} \)-constrained path and the 2D \( (Q_{20}, Q_{30}) \) surface gives insight into the onset of reflection-asymmetric shapes along the fission process.

The numerical calculations are performed using the \textsc{HFBTHO} code, which solves the HFB equations in an axially deformed harmonic-oscillator basis assuming axial symmetry \cite{Stoitsov2005CPC,Marevic2022CPC}. 
The Skyrme energy-density functional SLy4d \cite{Kim1997JPG} is employed to keep consistency with the dynamics calculations.
Pairing correlations are treated using a density-dependent delta interaction of mixed type
\begin{equation}
	V_{\text{pair}}(\bm{r}) = V_0 \left[ 1 - \eta \frac{\rho(\bm{r})}{\rho_0} \right] \delta(\bm{r}_1 - \bm{r}_2),
\end{equation}
where the pairing strengths $V_0=-360$ MeV for both neutrons and protons, with mixing factor $ \eta=0.5 $ and $\rho_0=0.16~{\rm fm}^{-3}$.

%

\subsection{TDHF calculations of quasifission dynamics}

The quasifission dynamics are investigated using the time-dependent Hartree-Fock (TDHF) theory with Skyrme interaction, which provides a fully microscopic description of the nuclear many-body evolution under a self-consistent mean-field. 
The time evolution of each single-particle wave function \( \varphi_i(\bm{r}, t) \) is governed by the TDHF equation
 \begin{equation}
	i\hbar \frac{\partial}{\partial t} \varphi_i(\bm{r}, t) = h \left[ \rho(t) \right]  \varphi_i(\bm{r}, t),
 \end{equation}
where \( h[\rho(t)] \) is the single-particle Hamiltonian derived from the instantaneous one-body density matrix \( \rho(t) \). 
The nucleon transfer, nuclear deformation, and energy dissipation are naturally incorporated through the self-consistent evolution of the mean-field in the TDHF theory.
More details on the TDHF formalism and its applications can be found in Refs.~\cite{Stevenson2019PPNP,Simenel2025EPJA}.

The numerical calculations are performed using the \textsc{Sky3D} code \cite{Abhishek2024CPC} with the Skyrme energy-density functional SLy4d \cite{Kim1997JPG}, which is fitted without center-of-mass corrections. 
All simulations are carried out on a $(24\times24\times48) \Delta d^3$ three-dimensional Cartesian grid with $\Delta d=1.0~\rm{fm}$, without assuming any spatial symmetry, and the time step $\Delta t=0.2$ fm/$c$. 
The initial conditions are generated by placing the projectile and target nuclei at a certain distance, each represented by its static HF-BCS ground state, and then imparting the appropriate boost corresponding to the desired center-of-mass energy $E_{\rm c.m.}$ and impact parameter.
During the time evolution, various observables such as the density distribution, quadrupole and octupole moment, and neck formation are monitored to characterize the reaction dynamics. 
The maximum simulation time for each TDHF event is set to 40 zs in this work.
An event is classified as a fusion if the dinuclear system remains unseparated at the end of this time window.

\section{\label{sec:results}Results and discussion}

%
%


\begin{table}[] 
	\centering
	\caption{\label{tab:channels} The reaction entrance channels calculated in this work.}
	\begin{ruledtabular}
		\begin{tabular}{ccc}
			\makecell{Entrance \\  channel} & \makecell{$E_{\rm c.m.}$ \\(MeV)} & \makecell{Orientation \\ (proj.-targ.)}  \\
			\hline
			$^{68}\mathrm{Zn}  + {^{112}\mathrm{Sn}}$	& 160.00 -- 167.00 	& tip-tip \\
			& 164.00 -- 171.60 	& side-tip\\
			
			$^{74}\mathrm{Se}  + {^{106}\mathrm{Pd}}$	& 166.00 -- 170.16 	& sph.-tip\\
			
			$^{80}\mathrm{Kr}  + {^{100}\mathrm{Ru}}$	& 168.00 -- 178.67 	& tip-tip\\
			& 163.00 - 177.22 	& side-tip\\

			$^{84}\mathrm{Kr}  + {^{96}\mathrm{Ru}}$	& 168.00 -- 182.74 	& side-tip  \\

		\end{tabular}
	\end{ruledtabular}
\end{table}

Table~\ref{tab:channels} lists the entrance channels calculated in this work.
Only the central collisions are considered in order to eliminate angular-momentum effects and enable a direct comparison with fission.
The $E_{\rm c.m.}$ of one entrance channel is increased with step sizes between 0.01 and 1.00 MeV in order to obtain a relatively successive evolution of the contact time.
By varying the entrance channels and the projectile orientations, the contact configuration is shifted to different locations in the deformation space, allowing us to probe a broad range of regions on the PES.

\begin{figure}
	\centering
	\includegraphics[width=8.6cm]{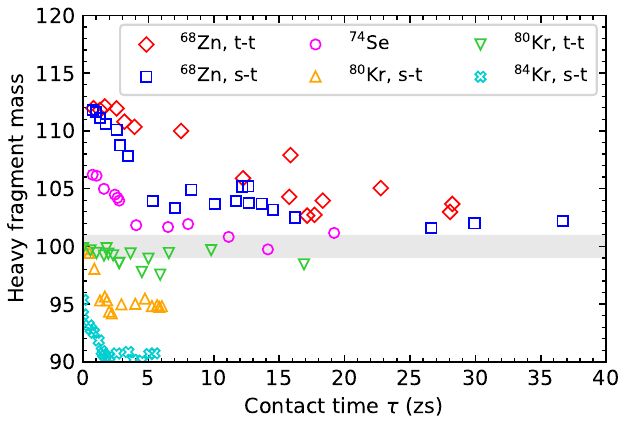}
	\caption{\label{fig:time_mass} The heavy-fragment mass as a function of the prefragments contact time $\tau$. The gray band represents the experimental data of the most probable heavy-fragment mass $100(1)$ in \Hg fission \cite{Andreyev2010PRL}. The letters ``s" and ``t"  denote the initial orientations ``side" and ``tip", respectively.}
\end{figure}

Figure \ref{fig:time_mass} shows the evolution of the heavy-fragment mass as a function of the contact time obtained from the TDHF calculations.
The contact time is defined as the duration during which the neck density of the dinuclear system remains above half of the nuclear saturation density $\rho_0/2\approx0.08~\mathrm{fm}^{-3}$.

Because of the initial mass asymmetry between the projectile and the target, the mass equilibration between fragments can be observed at all entrance channels. 
The decrease of the heavy-fragment mass represents the progress of this equilibration.
According to the liquid-drop picture, the surface energy of system reaches a minimum when the two fragments have equal sizes. 
Therefore, one generally expects the system to evolve toward complete mass equilibration as the contact time increases. 
Deviations from this trend indicate the influence of shell effects.

For the \ZnChan reaction, the mass equilibration proceeds with increasing contact time, but complete equilibration is not achieved even when the contact time exceeds the systematics-based estimate of $\approx 20$ zs \cite{Simenel2020PRL}. 
This behavior is driven by the shell effects.
The fusion hindered in \ZnChan reactions by shell effects is also observed in experiment \cite{Kozulin2021PLB}.
Notably, the equilibration process stops near a heavy-fragment mass of $A_{\rm H} = 100$, close to the fragment mass observed in the fission of $^{180}\rm{Hg}$.
A similar stagnation of the mass equilibration is also found in the \SeChan reaction. 
For the critical case where the entrance channel has the same mass as the fission fragments, namely the \KrChan system, the hindrance to equilibration appears only for the tip–tip orientation.
For \KrChanb with initial asymmetry less than $80/100$, no hindrance to mass equilibration is observed.

\begin{figure}
	\centering
	\includegraphics[width=9cm]{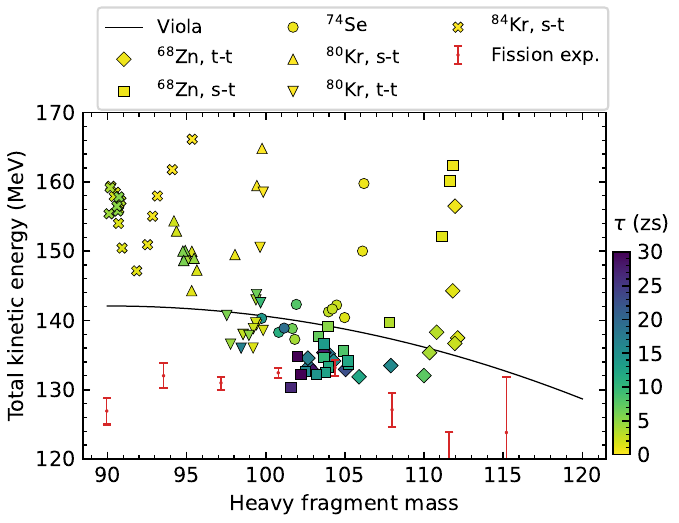}
	\caption{\label{fig:tke_mass} Total kinetic energy of fragments as a function of heavy-fragment mass. The letters ``s" and ``t"  denote the initial orientations ``side" and ``tip", respectively. The contact time $\tau$ for each reaction is indicated by the face color of each point. The black curve is given by Viola systematic \cite{Hinde1987NPA}. The experimental data of fission fragments TKE are taken from Ref.~\cite{Elseviers2013PRC}.}
\end{figure}

For strongly dissipative quasifission events, the total kinetic energy (TKE) is dominated by the Coulomb repulsion. 
Therefore, the TKE provides important information about the configuration at scission.
A more elongated configuration leads to a smaller Coulomb potential, whereas a compact configuration gives a larger Coulomb energy.
Figure \ref{fig:tke_mass} shows the fragment TKE as a function of the heavy-fragment mass for each event.
For strongly dissipative events, the calculated TKE is generally consistent with the Viola systematics \cite{Hinde1987NPA}.
Notably, the \ZnChan reaction yields TKE values that agree remarkably well with the experimental fission TKE of  $^{180}\rm{Hg}$, whereas the other  reactions produce relatively higher values.


\begin{figure*}
	\centering
	\includegraphics[width=\linewidth]{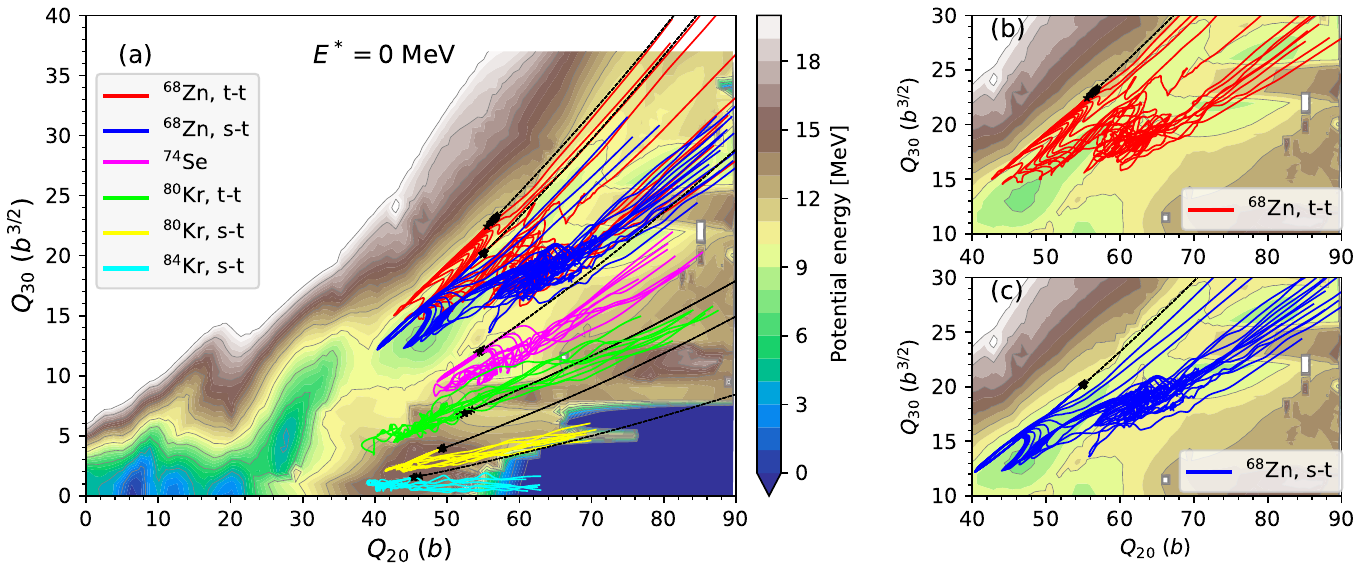}
	\caption{\label{fig:all_qf} (a) Quasifission trajectories of all entrance channels superimposed on the \Hg potential-energy surface. (b)(c) The enlarged views highlighting the trajectories of the \ZnChan reaction. The black dashed line and the colored solid line represent the entrance and the contact stage of reaction, respectively (see text for details). The stars indicate the contact point of the projectile and target. The letters ``s" and ``t"  denote the initial orientations ``side" and ``tip", respectively.}
\end{figure*}

The enhanced production of fragments with masses near $80/100$ in long–contact-time quasifission events, together with the similar TKE values, reflects characteristics analogous to those observed in the fission of $^{180}\rm{Hg}$.
To interpret the quasifission dynamics in terms of shell effects, the PES is used \cite{McGlynn2023PRC,Lee2024PRC,Simenel2025PLB}.
Figure~\ref{fig:all_qf} compares the PES with the reaction trajectories of each quasifission event in the $Q_{20}$–$Q_{30}$ space.
The incoming stage of the reaction is shown by black dashed lines.
Once the projectile and target make contact and the neck density exceeds $0.08~\mathrm{fm}^{-3}$, the trajectory is plotted as colored solid lines and terminates at scission, defined as the point where the neck density drops below $0.08~\mathrm{fm}^{-3}$.
Only events with contact times greater than $2~\mathrm{zs}$ \cite{Lee2024PRC}, which ensures sufficient kinetic-energy dissipation, are included in Fig.~\ref{fig:all_qf}.

For the \ZnChan reaction, the trajectories are guided by the PES valley located around $Q_{20}=40$--$60~\mathrm{b}$.
As the system approaches the second-saddle region near $Q_{20}=60~\mathrm{b}$, the diminishing depth of the valley weakens its guiding influence, causing the trajectory to exhibit noticeable wandering.
After crossing the saddle, the system continues to elongate until reaching scission.

The \SeChan reaction does not exhibit significant valley-driven evolution. 
This behavior is not attributed to the temperature effects on the PES, as detailed in Appendix~\ref{appendix:temperature}.
The observed hindrance of mass equilibration instead arises from the large PES gradients associated with the prominent ridge in the southern region of the trajectories.
A similar ridge effect is also present in the \KrChan reaction with tip–tip orientation.
The trajectories evolve along the edge of the ridge without ever fully crossing it.
In contrast, for the side–tip orientation, the trajectory remains in a region with much smaller PES gradient and thus does not exhibit the same hindrance to mass equilibration as in the tip–tip case.
The comparison between the two orientations of \KrChan clearly delineates the boundary of the region where the PES ridge exerts a decisive influence.

The influence of the symmetric valley at large elongation ($Q_{20} > 50~\mathrm{b}$) is demonstrated by the \KrChanb system.
The mass equilibration reaches completion at $Q_{20} \approx 40\text{ b}$, which is before the trajectories enter the symmetric valley region.
It indicates that mass equilibration in the $^{84}\text{Kr}$ systems is not primarily governed by the symmetric valley. This behavior is similar in the $^{80}\text{Kr}$ (side-tip) systems.
In contrast, the symmetric valley plays a critical role in limiting the dissipation of TKE. The potential gradient is large and therefore the system cannot undergo enough dissipation, leading to the V-shaped dependence of TKE on $E_{\text{c.m.}}$ observed for both the $^{80}\text{Kr}$ (side-tip) and $^{84}\text{Kr}$ systems in Fig.~\ref{fig:tke_mass}.
%
	

The trajectories reveal the role of the PES topography in quasifission fragment formation,
particularly that of the ridge, because the specific fragments enhanced by the ridge are difficult to directly infer from PES.
The valleys and ridges on the PES originate from shell effects.
Indeed, without shell corrections, a purely macroscopic liquid-drop potential would not exhibit various topographic features, resulting instead in a symmetric split.
Since quasifission is inherently a dynamical process, the non-equilibrium effects may contribute, such as collective inertia and collective friction. 
There is also the influence of excitation energy. 
Thus, the trajectories do not necessarily show an immediate or sensitive response to the fine structures of the zero-temperature PES, especially at the early stage of reaction.


\begin{table}[] 
	\centering
	\caption{\label{tab:compare} Properties of the three reactions selected for the detailed comparison.}
	\begin{ruledtabular}
		\begin{tabular}{ccccc}
			\makecell{Entrance \\  channel} & \makecell{$E_{\rm c.m.}$ \\(MeV)} & \makecell{Orientation \\ (proj.-targ.)}  & \makecell{Contact \\  time (zs)} & \makecell{TKE (MeV)}\\
			\hline
			$^{68}\mathrm{Zn}  + ^{112}\mathrm{Sn}$	& 171.56	& side-tip & 16.2 & 132.7  \\
			
			$^{74}\mathrm{Se}  + ^{106}\mathrm{Pd}$	& 170.16 	& sph.-tip & 19.2 & 138.9 \\
			
			$^{80}\mathrm{Kr}  + ^{100}\mathrm{Ru}$	& 177.22 	& tip-tip & 16.9 & 136.0 \\
			
		\end{tabular}
	\end{ruledtabular}
\end{table}
\begin{figure}
	\centering
	\includegraphics[width=9cm]{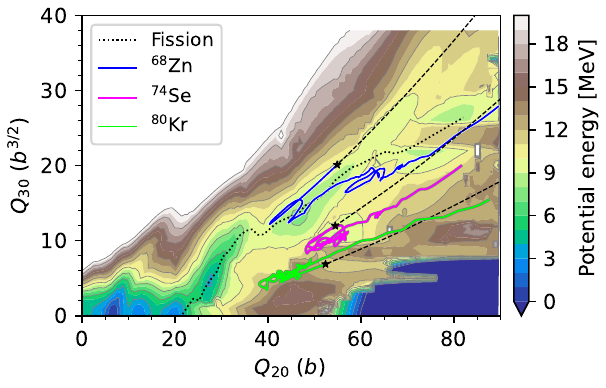}
	\caption{The trajectories of quasifission listed in Table~\ref{tab:compare} and the static \Hg fission path on top of \Hg potential-energy surface. The stars indicate the contact point of the projectile and target.}
	\label{fig:3traj}
\end{figure}

Although the quasifissions can enhance the formation of $80/100$ mass fragments, it remains unclear whether all these fragments can be considered similar to the \Hg fission one, given the TKE differences observed in Fig.~\ref{fig:tke_mass}.
To address this, we compare three quasifission events listed in Table~\ref{tab:compare} and the static fission path.
The static fission path is obtained from $ Q_{20} $-constrained HFB calculations.
These quasifission events are selected based on contact time comparable to the longest one for \KrChan (tip-tip), ensuring a similar degree of kinetic-energy dissipation and reducing the influence of variations in the contact time.

The trajectories for the three quasifission events, together with the fission path, are shown in Fig.~\ref{fig:3traj}.
The static fission path agrees with other microscopic calculations \cite{Zeyu2022PRC,Bernard2024EPJA} and gives a light-fragment mass of $81.5$ at the scission point, which is close to the experimental value of $80(1)$.
The \ZnChan trajectory is close to the fission path, suggesting that a similar mechanism may be involved.

\begin{figure}
	\centering
	\includegraphics[width=8.6cm]{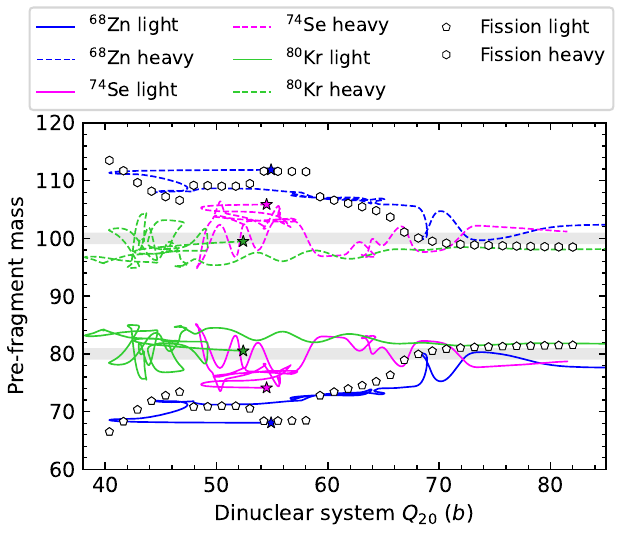}
	\caption{The prefragment masses as a function of the dinuclear system $Q_{20}$ for the quasifission events listed in Table~\ref{tab:compare} and for the fission of $^{180}\rm{Hg}$. The stars indicate the contact points of the projectile and target for each quasifission. The gray bands indicate the mass regions around $80\pm1$ and $100\pm1$.}
	\label{fig:masssq20}
\end{figure}

We define, as follows, a neck plane to separate the dinuclear system in order to analyze the mass–equilibration process between the prefragments within a single reaction event.
The eigenvector of quadrupole tensor corresponding to the largest $Q_{20}$ is first taken as the main axis of rotation of the dinuclear system.  
The nuclear density is then projected onto this axis, and the minimum of the one-dimensional density distribution between two maxima is taken as the neck plane that divides the system into two prefragments.  
Figure~\ref{fig:masssq20} presents the evolution of the prefragment masses as a function of the dinuclear system's $Q_{20}$. 
The pentagons and hexagons are obtained from the static fission path calculations.
The results show that, under the guiding effect of the potential valley, the \ZnChan quasifission exhibits a very small mass shift of less than two nucleons before crossing the second saddle ($Q_{20} \approx 60~\mathrm{b}$).
The equilibration occurs only after the saddle and then freezes again near the  $80/100$ mass split.
This behavior aligns with the result in microscopic static fission calculation.
In contrast, for the \SeChan and \KrChan cases, the equilibration hindrance appears only near the $80/100$ mass split, consistent with the PES topology where the potential ridge is the only common topographic feature.

\begin{figure}
	\centering
	\includegraphics[width=8.2cm]{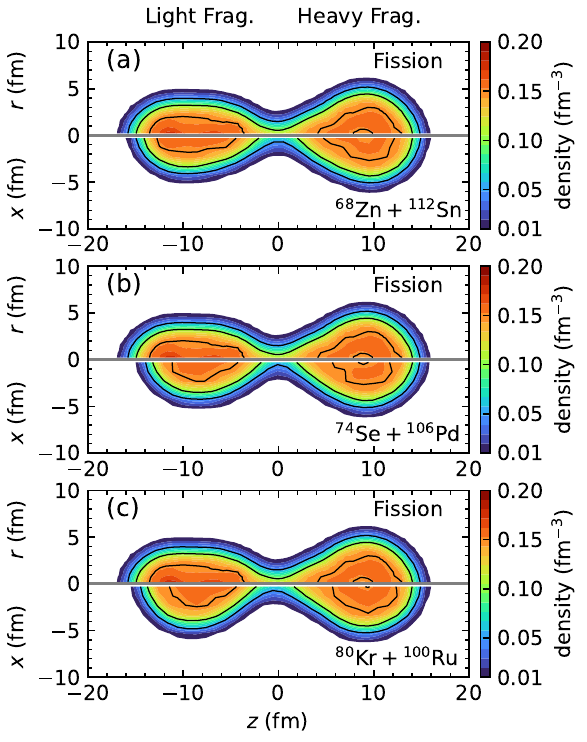}
	\caption{Comparison of the nuclear shapes at the scission point from the TDHF quasifission calculations (lower panel) and the static \Hg fission calculation (upper panel).
		The density contours correspond to 0.05, 0.10, and 0.15 $\mathrm{fm}^{-3}$.
		The quasifissions shown are those listed in Table~\ref{tab:compare}.}
	\label{fig:scission_density}
\end{figure}

Figure~\ref{fig:scission_density} compares the densities at scission point of the three quasifission reactions with that of \Hg fission.
The heavy fragments in all three quasifission cases exhibit configurations very similar to the heavy fragment in fission.
In contrast, for the light fragment, only the \ZnChan quasifission shows a configuration comparable to that in fission, characterized by a more elongated shape.
Such a more elongated configuration explains the smaller TKE obtained in the \ZnChan calculation.
The \SeChan and \KrChan quasifissions instead produce more compact light fragments.

Compared with the \SeChan and \KrChan reactions, the TDHF calculation of the \ZnChan quasifission shows consistency with the microscopic static calculation of \Hg fission, and particularly yields TKE in good agreement with the experimental fission TKE.
The TKE predicted by TDHF is independent of any artificial definition of the scission point.
And TDHF has been shown to reliably reproduce dissipation mechanisms in low-energy reactions \cite{Williams2018PRL},
where the two-body dissipation is found to be insignificant \cite{Huang2025PLB_role}.
The TKE of \ZnChan support the implication that in forming \Hg fission-like fragments, the potential valley along the asymmetric path still plays an important role, rather than the dynamics being driven solely by the hindrance of the potential ridge.
In addition, the differences in the calculated TKE arise from the elongation of the light fragment, suggesting that such deformation is a crucial factor in the experimentally observed proton-shell stabilization of light fragments in preactinide fission \cite{Mahata2022PLB,Buete2025PLB,Morfouace2025Nature}.

\begin{figure*}
	\centering
	\includegraphics[width=\linewidth]{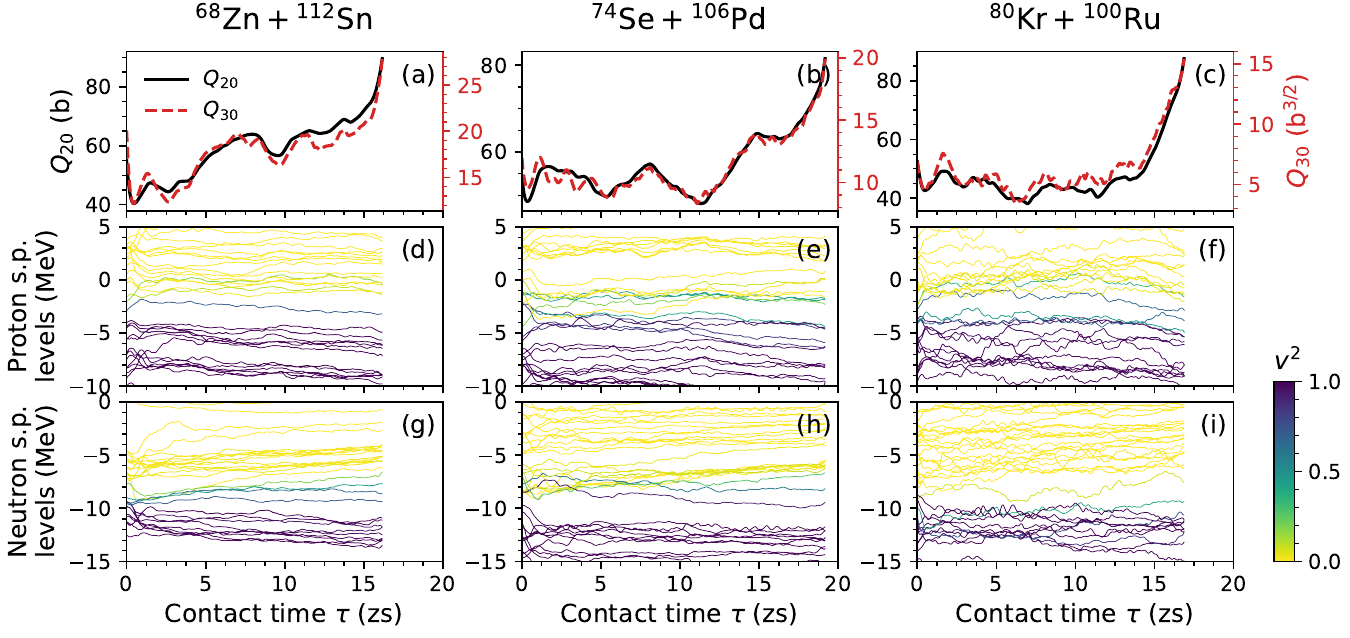}
	\caption{
		Time evolution of (a)--(c) the multipole moments $Q_{20}$ and $Q_{30}$ of the compound system, and (d)--(i) the single-particle levels for the reactions in Table~\ref{tab:compare}. 
		The color mapping indicates the occupations $v^2$.}
	\label{fig:sp_levles}
\end{figure*}

To reveal the role of proton and neutron shell effects, Fig.~\ref{fig:sp_levles} shows the time evolution of proton and neutron single-particle levels for the three quasifissions.
Due to the frozen-occupations approximation, occupations near the Fermi surface remain static,
though typically fewer than five levels are involved.
For the \ZnChan reaction, a robust proton shell gap ($\approx 3 \text{ MeV}$) forms early and remains stable, whereas an 1 MeV weaker neutron shell gap emerges during the early stage ($\tau < 5$ zs) but does not persist. 
For the \SeChan and \KrChan systems, the proton-shell gaps are approximately $1\text{--}2$ MeV narrower than those of $^{68}\text{Zn} + ^{112}\text{Sn}$, while the neutron gaps are about $2$ MeV wider.
This prominent neutron shell likely contributes to preventing the \SeChan trajectory from entering the potential valley on the asymmetric fission path.
A similar neutron-shell gap is also observed in the side-tip \KrChan system. 
Therefore, this neutron shell gap does not account for the hindrance to mass equilibration.
These results support the important role of the proton shell effects in \Hg fission. 
Further systematic calculations will be interesting, to verify the stability of proton shell effects across the preactinide region \cite{Mahata2022PLB,Buete2025PLB,Morfouace2025Nature}.



\section{\label{sec:summary}Summary}

In this work, the TDHF approach is employed to investigate the quasifissions in central collisions of \ZnChan, \SeChan, \KrChan, and \KrChanb at several incident energies.
These reaction systems all correspond to the same compound nucleus ($^{180}\mathrm{Hg}$).
We observe shell–driven hindrance of mass equilibration between the prefragments, leading to an enhanced production of fragments near the $80/100$ mass split, similar to that seen in \Hg fission.

A comparison between the quasifission trajectories and the PES in $(Q_{20}, Q_{30})$ space demonstrates the key role of PES topography,
in particular, the role of the potential ridge in forming $80/100$ mass fragments.
This underscores the need for dynamical calculations.
In contrast to actinides, where deep potential valleys allow a relatively intuitive interpretation of fragment formation, the PES topography in preactinides is more subtle and cannot be reliably understood without dynamical evolution.

Compared with the \SeChan and \KrChan reactions, the \ZnChan quasifission exhibits a prefragment equilibration process and a scission configuration that align with those of the static fission calculation.
In addition, the TDHF-predicted TKE of \ZnChan quasifission shows good agreement with the experimental TKE of \Hg fission.
These suggest that, although the potential ridge alone can enhance the formation of $80/100$ mass fragments, the potential valley on the asymmetric path plays an essential role in producing \Hg fission-like fragments.
The differences in TKE among the quasifissions originate from the deformation of the light fragment.
Meanwhile, we show the presence of proton shell gap in the dinuclear system of \ZnChan quasifission.
An accurate description of the light-fragment shape may therefore be essential for understanding the experimentally observed proton-number stabilization of the light fragment in the systematics of the preactinide fission.

\section*{ACKNOWLEDGMENTS}

We thank C. Simenel for inspiring discussions. 
We also thank Zepeng Gao, Sibo Wang, Yinu Zhang, Yu Yang, and Fuchang Gu for valuable discussions. 
This work is supported by the Guangdong Basic and Applied Basic Research Foundation No. 2024A1515012310, the National Natural Science Foundation of China under Grant No. 12475136, and the China Scholarship Council under State Scholarship Fund No. 202406380014.

\appendix
\section{Finite-temperature effects on PES}\label{appendix:temperature}
\begin{figure*}
	\centering
	\includegraphics[width=\linewidth]{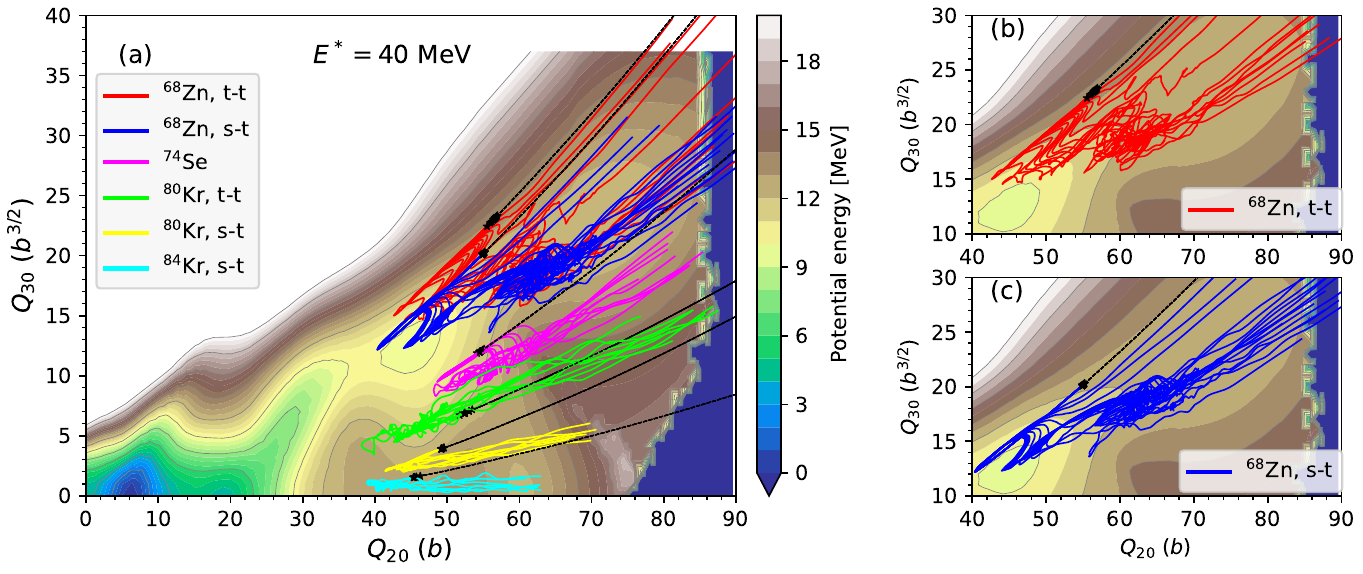}
	\caption{Same as the Fig.~\ref{fig:all_qf}, but on the finite-temperature PES at $E^*=40$ MeV.}
	\label{fig:all_qf_e40}
\end{figure*}

To evaluate the weakening of shell and pairing effects as temperature increases, the finite-temperature HFB (FT-HFB) formalism is applied \cite{Stoitsov2013CPC_axially,Schunck2012CPC_solution}.
The FT-HFB equations are formally identical to the HFB ones at temperature $T = 0$, with the distinction that the density matrix and pairing tensor depend on the Fermi–Dirac occupations $f_\mu$ of the quasiparticle states $\mu$
\begin{equation}
	f_\mu = \frac{1}{1 + \exp(E_\mu / k_B T)},
\end{equation}
where $E_\mu$ denotes the quasiparticle energy and $k_B$ is the Boltzmann constant.
$T$ is calculated using $T = \sqrt{E^*/a}$ with $a = A/10$.

In TDHF calculations, $E^*$ can be determined by the difference between $E_{\text{c.m.}}$ and the TKE.
For the reactions listed in Table~\ref{tab:channels}, the maximum $E^*$ is estimated to be approximately $40~\text{MeV}$.
This value is consistent with estimates derived from the Coulomb barrier ($V_\mathrm{b} \approx 175~\mathrm{MeV}$) and the reaction $Q$-value ($ \approx -140~\mathrm{MeV}$), and corresponds to a nuclear temperature of $T = 1.49~\mathrm{MeV}$.
FT-HFB calculations are performed to evaluate the PES at this temperature, as shown in Fig.~\ref{fig:all_qf_e40}. 
While details deviate from the zero-temperature PES, the primary topological features—namely the asymmetric path valley and the potential ridge—persist.
It is consistent with the results that the \ZnChan reactions remains governed by the influence of the asymmetric valley.

It can be noted that the apparent disappearance of the symmetric valley at large elongation ($Q_{20} > 50~\mathrm{b}$) in the finite-temperature PES.
We attribute that to discontinuities encountered near the scission point, where the system undergoes significant configuration transitions \cite{Dubray2012CPC_numerical,Carpentier2024PRL_construction}. 
The scission points of $^{84}\text{Kr} + ^{96}\text{Ru}$ trajectories support the persistence of a deep symmetric valley within this region.



\bibliography{apssamp_prc}

\end{document}